# Open Source Physics

*By Wee Loo Kang*

## What

Open Source Physics (Brown, 2012; Christian, 2010; Esquembre, 2012; Hwang, 2010) empowers teachers and students to create and use these free tools with the associated intellectual property rights given to customise (Wee & Mak, 2009) the computer models/tools to suit their teaching and learning needs.

Open Source Physics (OSP) focuses on design of computer models, such as Easy Java Simulations (EJS) and the use of video modeling and analysis (Tracker). They allow students to **investigate, explore and analyse** data which is either real or simulated.

### Technologies

Easy Java Simulation
http://www.um.es/fem/EjsWiki/pmwiki.php
Tracker
http://www.cabrillo.edu/~dbrown/tracker/

Runs on Windows, MacOSX and Linux
Java Runtime  http://java.com/en/download/

The OSP approach helps users overcome barriers in creating, using and scaling up meaningful ICT use in education.  In Singapore, teachers and students have created or customised existing computer models to design and re-purpose EJS models to suit their context and learning needs.

Tracker tools allow students to analyse different aspects of a physics phenomena to deepen their understanding of abstract physics concepts.

Using Tracker, students record the motion of objects and create their own models to test their ideas in the study of kinematics.

Essentially, the two main pedagogical approaches (Figure 1) used in OSP are:

- **Guided Inquiry**
   o investigating modeled phenomena(EJS)
   o video analysis (Tracker)
- **Constructionism**
   o learning by making new models(EJS)
   o video modeling(Tracker)

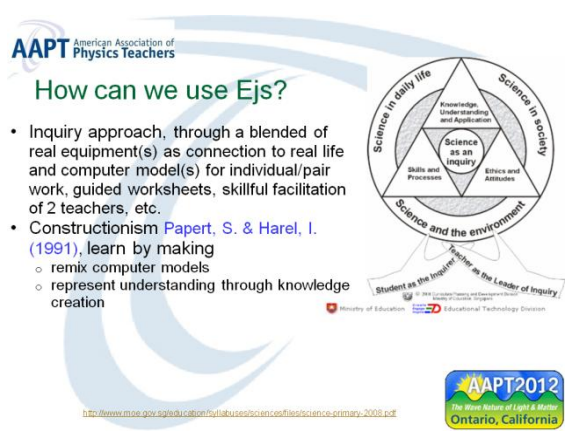

Figure 1. Taken from: Wee, L. K. (2012, Feb 4-8). AAPT 2012 Conference Presentation:Physics Educators as Designers of Simulations. *2012 AAPT Winter Meeting*

## Why

Brown Professor of Physics of Davision College, Wolfang Christian[1] calls for learners to be like scientists and "to come up with their own questions that they want answered".

The Easy Java Simulation authoring toolkit can be used to create computer models either by teachers to help students learn; or by students to represent their understanding.

---

[1] Christian, W. (2012). Building a National Digital Library for Computational Physics Education Podcast, from http://vimeo.com/37964196

*Open source licensing of educational computer models/software has opened up tremendous opportunities for learning in the 21$^{st}$ century, since anyone can freely access and modify these resources to suit their teaching and learning needs.*

The use of **OSP tools builds and deepens teachers' professionalism**[2] as they become designers of learning environments through customisation of existing computer models in the Open Source Physics comPADRE Digital Library Collection.

## Awards

MOE Innergy Award Winner Gold 2012 (Gravity-Physics by Inquiry)
MOE Innergy Award Winner School Commendation 2011 (Learning Physics through video analysis RVHS)

## Self-Directed Learning

Ownership of learning
- Students set learning goals and identify learning tasks to achieve the goals

Management and monitoring of own learning
- Students explore alternatives and make sound decisions
- Students formulate questions and generate own inquiries
- Students plan and manage workload and time effectively and efficiently
- Students reflect on their learning and use feedback to improve their schoolwork

Extension of own learning
- Students apply learning in new contexts
- Students learn beyond the curriculum

## Collaborative Learning

Effective group processes
- Students interactively contribute own ideas clearly and consider other points of view objectively and maturely
- Students ask questions to clarify and offer constructive feedback

Individual and group accountability of learning
- Students rely on each other for success

## How

Students are taught how to use the EJS models (Figure 2) and Tracker tool, so that they can conduct the learning task more effectively.

Each guided inquiry activity with EJS or Tracker usually lasts between 1 and 1.5 hours. Advanced learners can become more deeply engaged in project-based learning, for 6 to 10 weeks, conducting research on a problem which is personally motivating.

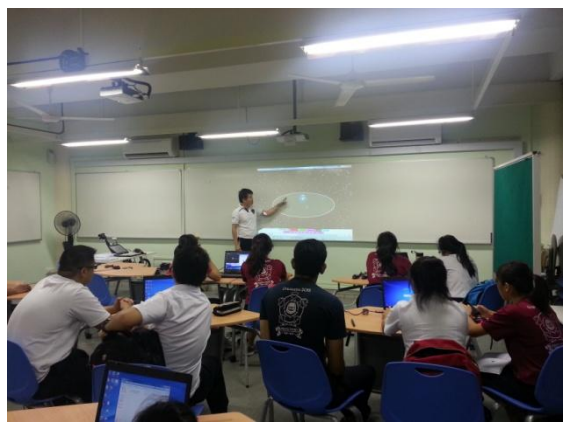

Figure 2. Teacher, Jimmy Goh of Yishun Junior College explaining and demonstrating affordances of EJS model and inquiry tasks. Photo by Wee Loo Kang

---

[2] Teachers — The Heart of Quality Education Retrieved 20 October, 2010, from http://www.moe.gov.sg/media/press/2009/09/teachers-the-heart-of-quality.php

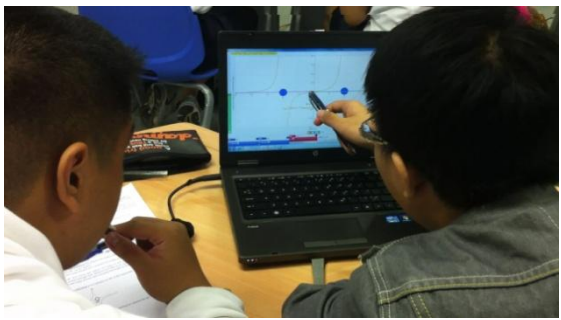

Figure 3. Students in pairs conducting experiments on EJS model and taking notes of their findings. Photo by Wee Loo Kang

Students typically work in pairs (Figure 3) to investigate a physical phenomenon using the technology. They come up with their own 'what-if' scenarios and **simulate them to discover and better understand the physics behind it**. Peer instruction is another way by which students make sense of the data. The tool is designed to support inquiry and the inquiry nature of the learning allows students to work and think like scientists.

Finally, students may be invited to present their analysis and defend their interpretation of the findings before the entire class.

A budding community of physics teachers, led by Wee Loo Kang [3] from the Educational Technology Division, has remixed 75 EJS models and lessons packages and numerous tracker video lessons, with the vision of developing a national digital library in Singapore for the benefit of anyone worldwide.

---

[3] For his work on Open Source Physics, Wee Loo Kang was awarded the MOE Outstanding Innovator Award 2013 and Public Service PS21 Excel Awards Best Ideator 2012.

## ETD Projects

**Using EJS**
eduLab Project
River Valley High School
Yishun JC
Anderson JC
Innova JC
Serangoon JC

**Using Tracker**
Proposed eduLab Project
Raffles Girl's School
River Valley High School
Greenridge Secondary
National JC
Evergreen Secondary

## Resources

Brown, D. (2012). Tracker Free Video Analysis and Modeling Tool for Physics Education, from http://www.cabrillo.edu/~dbrown/tracker/

Christian, W. (2010). Open Source Physics (OSP) Retrieved 25 August, 2010, from http://www.compadre.org/osp/
Esquembre, F. (2012). Easy Java Simulations Retrieved 13 September, 2012, from http://www.um.es/fem/EjsWiki/pmwiki.php

Hwang, F.-K. (2010). NTNU Virtual Physics Laboratory Retrieved 13 September, 2012, from http://www.phy.ntnu.edu.tw/ntnujava/index.php & http://www.phy.ntnu.edu.tw/ntnujava/index.php?board=23.0

Wee, L. K. (2012, Feb 4-8). AAPT 2012 Conference Presentation:Physics Educators as Designers of Simulations. 2012 AAPT Winter Meeting, from http://arxiv.org/ftp/arxiv/papers/1211/1211.1118.pdf

Wee, L. K., & Mak, W. K. (2009, 02 June). Leveraging on Easy Java Simulation tool and open source computer simulation library to create interactive digital media for mass customization of high school physics curriculum. Paper presented at the 3rd Redesigning Pedagogy International Conference, Singapore.

## Participation

Join our Open Source Physics Learning Community. Teachers will have access to lesson resources, community support, and consultancy.

For more information contact
Wee_Loo_Kang@moe.gov.sg
Lye_Sze_Yee@moe.gov.sg

Wee L.K. (2013) Open Source Physics, i in Practice 1(1), p. 58-63, Ministry of Education [PDF] [Full PDF]

**Lesson Exemplar**

**Science Investigation in Daily Life: Learning Physics of Sport Science through Video Analysis & Modeling (Tracker)**

Level: Upper Sec to JC 1
Subject: Physics

Assessment indicators
*Evidence of learning is shown when students are able to:*

| | | |
|---|---|---|
| • Conduct a video analysis and modeling of the physics of a sport science of their choice 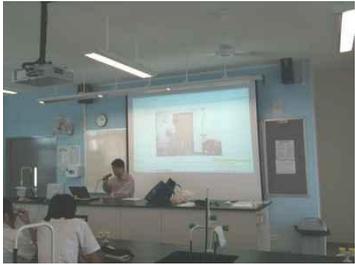 | • Select a suitable scientific question and setup to investigate on 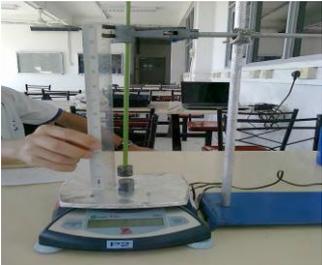 | • Create a scientific oriented analysis and model to substantiate a scientific report 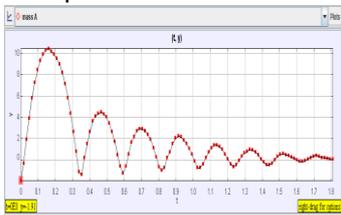 |

**LESSON ACTIVITY**

1. At the start of the activity, teacher may discuss steps in a scientific investigation by getting students to share previous scientific investigations carried out in primary science lessons.

2. Teachers are enouraged to introduce students to the pedagogy of video analysis and modeling. Students need to experience the sample analysis and modeling on one simple and one more complex video analysis and modeling. This is important because being a science literate citizen requires first-person experiences to practice these skills and processes as part of the science investigation.

    - <u>Pose questions</u>: This is the skill of asking suitable questions to initiate an investigation, in this case, a sport science that student find personally motivating.

    - <u>Plan investigation</u>: This is the process of devising ways to find the solution to the problem. This involves deciding on the measurements needed (displacement, velocity, acceleration, energies etc) and the types of equipment (video taken by the students, or otherwise) used to make these measurements.

    - <u>Conduct an investigation</u>: This is the process of carrying out the procedure to collect these measurements and observations. The measurements and observations are presented in a suitable way to report the findings to others.

    - <u>Analyse and evaluate results</u>: This is the skill of drawing conclusion from the investigation, and assessing whether the solution works, what is the physics being investigated etc.

    - <u>Communicate results</u>: This is the skill of presenting the conclusion to others.

3. Students are to first individually read through the problems from the other groups and then provide feedback to improve the latter's analysis.

4. Teacher can check with student-groups if they have faced the same problem and mentor them towards a scientific report.
5. Student-groups to present the scientific reports following a peer evaluation, contributing to a year-end formal assessment grade.

Teacher to mentor, grade and feedback on the areas of improvement for students to self evaluate what can be improved on and for follow-up action by the students-groups.

**Pedagogical and Assessment Considerations**

**Context**
- This activity aims to take students through the process of a scientific investigation, using sports science, a real life example of long jump, billiard balls motion, butterfly swimming stroke, basketball throw, rotating fan etc.

**Essential Takeaways**
- Teacher may share that Science is not just a collection of facts or ideas about things around them. It is a way of thinking and finding out about the physical and natural world. This project-based lesson uses tools that real physicists use for physics education.
- Students need to think of their own questions because the problems of tomorrow will require solutions that go beyond repeating what they already know today.
- It is important that we remain open-minded about the conclusions of the scientific investigation. We should anot be deterred if the outcomes of the investigation may not be what we have expected. Instead, we should reflect, evaluate and repeat the experiment again, if necessary.

**Supporting inquiry**
- Collecting and recording evidence are part of carrying out an investigation. Close mentoring is required to guide student-groups.

**Differentiating Instruction**
For higher ability students, teachers may guide the modeling construction to validate students' analysis and discuss the assumptions made in the investigation as well as provide suggestions for refinements to make the investigation better.

**Reference:** (Prepared by http://weelookang.blogspot.sg/)
http://ictconnection.opal.moe.edu.sg/cos/o.x?ptid=711&c=/ictconnection/ictlib&func=view&rid=82

*Lesson Exemplar from Mr Lee Tat Leong, River Valley High School*